\documentclass[prb,preprintnumbers,amsmath,amssymb,floatfix,aps,twocolumn]{revtex4}
\usepackage{graphicx}

\usepackage{xcolor}

\usepackage{hhline}
\usepackage{hyperref}

\newcommand\doilink[1]{\href{http://dx.doi.org/#1}{#1}}
\newcommand\doi[1]{DOI:\doilink{#1}}

\providecommand*\url[1]{\href{#1}{#1}}
\renewcommand*\url[1]{\href{#1}{\texttt{#1}}}

\begin{document}

\title{Thermopower of Graphene Nanoribbons in the Pseudodiffusive Regime}

\date{\today} 

\author{J. W. Gonz\'alez,$^{1,\,}$\footnote{Corresponding author: sgkgosaj@ehu.eus}, 
L. Rosales$^{2}$, A. Ayuela$^{1}$}
\affiliation{
(1) Centro de F\'{i}sica de Materiales (CSIC-UPV/EHU)-Material Physics Center (MPC), 
Donostia International Physics Center (DIPC), Departamento de F\'{i}sica de 
Materiales, Fac. Qu\'{i}micas UPV/EHU. Paseo Manuel de Lardizabal 5, 20018, 
San Sebasti\'an-Spain.\\
(2) Departamento de F\'{i}sica, Universidad T\'{e}cnica Federico Santa Mar\'{i}a, 
Casilla Postal 110V, Valpara\'{i}so, Chile.}

\begin{abstract}
Thermoelectric measurements for graphene ribbons are currently performed on samples that include atomic disorder via defects and irregular edges. In this work, we investigate the thermopower or Seebeck coefficient of graphene ribbons within the linear response theory and the Landauer formalism, and we consider the diffusive regime taken as a limit of the ribbon aspect ratio. We find that the thermopower and the electronic conductivity depend not only on the aspect ratio, but also on chemical potential and temperature, which are set here as key parameters. The obtained numerical results with temperature and doping are brought into contact with the thermoelectric measurements on disordered graphene ribbons with good agreement.
\end{abstract}

\maketitle

\section{\label{sec:intro} Introduction}
Graphene has become a main subject of research due its outstanding mechanical,
thermal and electronic properties, among which the high electron mobility 
stands out. The increased carrier mobility and the long mean free path at 
room temperature establish graphene as a good building material to fabricate
microwave transistors, photodetectors, and other electronic 
devices.\cite{tassin2013graphene,gonzalez2011transport}
Most of the proposed graphene devices consider ballistic electron propagation, 
so that the mean free path is longer than the size of the device, and electrons
move across the sample without undergoing inelastic scatterings that break 
phase coherence. Although the ballistic transport regime has been studied both theoretically and experimentally,\cite{miao2007phase} graphene has several types of defects, known to act as scattering centers within nanostructures that could modify their typical electronic properties.

Understanding the electron mobility of graphene samples requires studying 
the role of impurities and defects. Experiments on disordered 
two-dimensional graphene show a residual
conductivity $4e^2/\pi h$ that arises from the electron density of 
states not vanishing  at zero gate voltage.\cite{tan2007measurement,chen2008charged,martin2008observation,xia2009measurement}
In agreement with the experiments, calculations using the semi-classical 
Boltzmann equation or the quantum Kubo formula for electron transmission show 
that dirty samples with a large concentration of charged impurities have a 
minimum conductivity value at low carrier density of about $4e^2/\pi h$ even 
for ideal pristine graphene.\cite{adam2007self,nomura2007quantum,tworzydlo2006sub}
%
%
Furthermore, calculations using the Green-Kubo theory in large length ribbons 
having randomly distributed disorder, such as single and double vacancies, 
Stone-Wales defects and irregular ribbon-edge terminations, show that the phonon 
contribution to thermal conductance is negligible compared to the electron 
contribution.\cite{haskins2011control}

Thermopower in the diffusive limit is measured in large and defective 
samples of exfoliated graphene
and CVD-graphene.\cite{zuev2009thermoelectric,wei2009anomalous,babichev2013resistivity,xiao2011enhanced} 
The experimental thermopower measurements show a linear dependence with 
temperature, following the semi-classical Mott formula, which is also based 
on a weak electron-phonon interaction and a negligible phonon-drag effect.\cite{sankeshwar2013thermoelectric}
The questions that have to be answered next are: how the thermopower changes when the electron transport passes from the ballistic regime to the diffusive one, 
and how to characterize the diffusive limit in the thermopower calculations of long graphene samples.
We herein study these questions using numerical and analytical 
approaches,\cite{borunda2013ballistic,tworzydlo2006sub,cresti2013impact} and we find that when the conductivity enters the pseudodiffusive regime, the thermopower tends to be a constant directly proportional to temperature.

\begin{figure}[!ht]
\includegraphics[clip,width=0.45\textwidth,angle=0,clip]{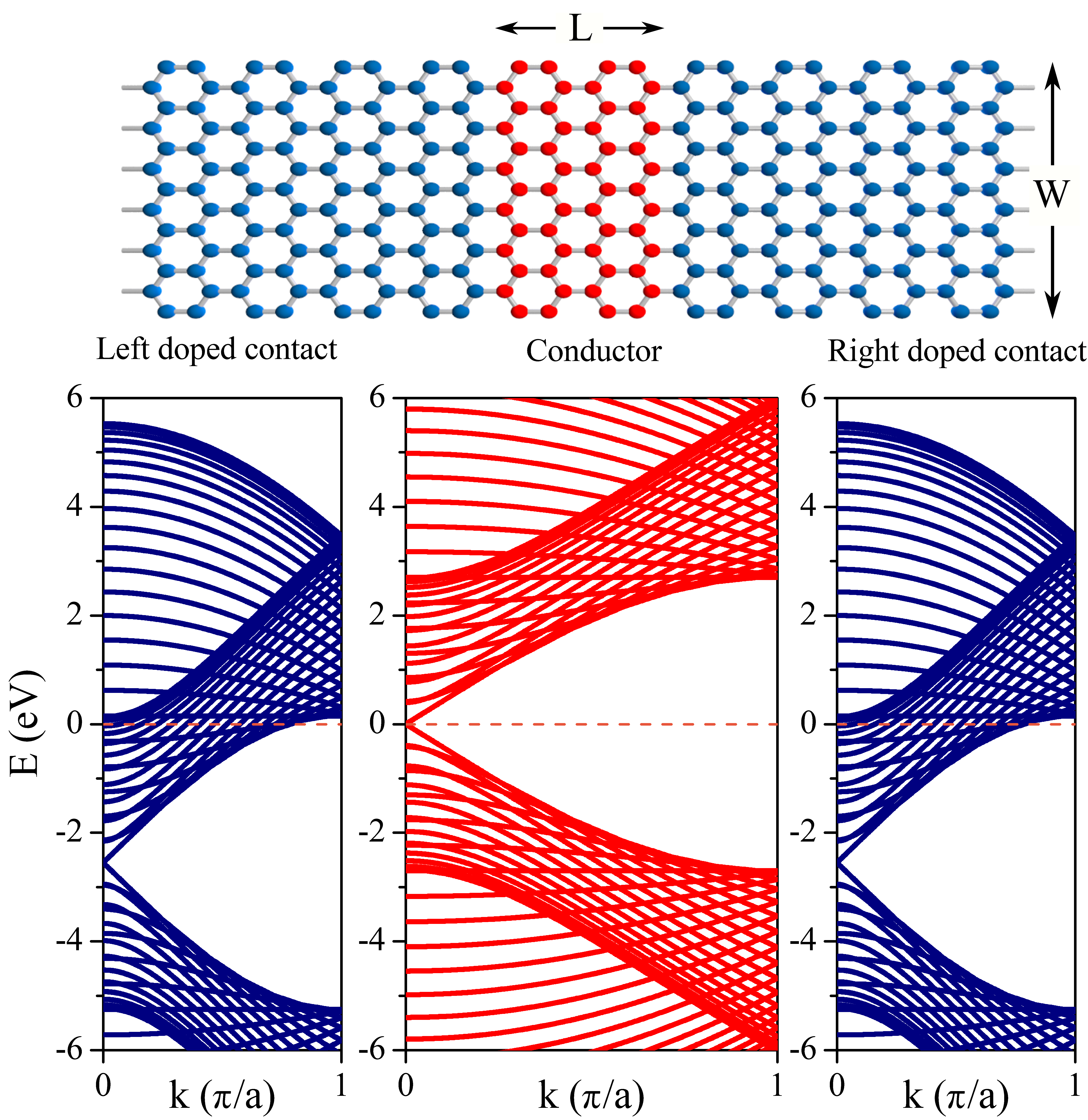} 
\caption{(Color on-line) Band alignment of the conductor and the n-doped leads.
We choose a metallic armchair graphene nanoribbon $N = 35$ $\left(W = 35 \sqrt{3} a_{CC}\right)$ and $V_{gate}^{lead} = -2.65$ eV to maximize the number of available conducting channels in the leads. 
}
\label{Fig:Band_35LCR}
\end{figure}

%
%
%

\section{\label{sec:theory} Model}
The diffusive regime is reached for highly disordered samples when the 
average distance between scattering centers becomes larger than the typical 
coherence length $L_\phi$.\cite{dietl2009disorder} Since the graphene 
experiments estimate the coherence length between 3 and 5 $\mu$m at $260$ mK.\cite{miao2007phase}, atomistic models in the diffusive regime should involve 
a large number of atoms.  
Instead, without including the disorder explicitly, we consider the so-called ``pseudodiffusive'' regime defined when the transmission coefficients of the pristine system in a given configuration behaves as in a diffuse regime, 
even without the presence of disorder.
This pseudo-diffusive regime can be achieved in several configurations, via electronic confinement through quantum dots\cite{borunda2013ballistic} and in  graphene stripes being wide and short.\cite{tworzydlo2006sub,cresti2013impact} 
We herein choose the latter option and study pristine metallic armchair 
nanoribbon of width $W$ in a tunnel junction configuration, with highly 
doped contacts and an undoped conductor section of length $L$. The leads 
are taken as the same nanoribbon doped using a large on-site potential, 
as shown in the model of Fig. \ref{Fig:Band_35LCR}.

\subsection{\label{sec:numerical} Numerical Approach}
We first focus on the electronic transport properties calculated using the Landauer formula and the Green's function matching 
formalism.\cite{nardelli1999electronic,datta1997electronic} 
This method divides the system into three blocks. The central region 
described by the Hamiltonian matrix $\mathcal{H}_\text{C}$ has a finite size and is geometrically parametrized using the width $W$ and the length $L$.
The low energy transport properties of the aGNRs are well described using a 
$\pi$-band tight-binding approximation with nearest-neighbor hopping.\cite{gonzalez2011transport,nardelli1999electronic,gonzalez2015electron} 
The  Hamiltonian of the central region can be written as
\begin{equation}
\mathcal{H}_\text{C} = \sum_{m}{\varepsilon_\text{C}}c_m^\dag {c_m} 
   - \gamma_0\sum_{\langle i,j\rangle} c_i^\dag c_j \,,
 \label{ham:FS}
\end{equation}
where $\gamma _0=2.75$ eV is the hopping energy between nearest-neighbor 
carbon atoms,\cite{Saito:2003,Charlier:2007,Laird:2015} $c_m ( c_m^\dag )$ 
is the annihilation (creation) operator on the $m$-th site of graphene 
lattice, and $\varepsilon_\text{C}$ is the carbon on-site energy controlling 
the n/p electronic dopping.\cite{gonzalez2010bound}
The contacts are labelled on the left (L) and on the right (R) are two 
semi-infinite leads made of the same pristine metallic armchair graphene 
nanorribons, and are described with Hamiltonians $\mathcal{H}_\text{L(R)}$. 
The doped leads are described using a large potential, $V_{gate}^{lead} = -2.65$ eV to maximize the number of available conducting channels.

The transmission coefficients in the linear response approach are calculated
within the Green's function formalism given by 
\begin{equation}\label{LandauerG}
\mathcal{T}\left( {E } \right) =
{\mathop{\rm Tr}\nolimits} \left[ {\Gamma_L G_C^R \Gamma _R G_C^A }\,\,
\right],
\end{equation}
where the retarded (advanced) conductor Green's function, $G_C^{R(A)}$ is 
obtained following  a real-space renormalization  scheme, and $\Gamma_{L,R} $
describes the particle scattering between the R,L lead and the 
conductor.\cite{nardelli1999electronic,gonzalez2010bound,gonzalez2011transport,gonzalez2015electron} 
The electronic conductance is thus defined as 
$G\left( {E } \right) = \frac{2e^2}{h}\mathcal{T}\left( {E } \right)$ 
and the conductivity $\sigma\left( {E } \right) = G\left( {E } \right) \times L/W$.

We then investigate the thermopower or Seebeck coefficient $S$. 
The thermopower is defined as the voltage drop induced by the temperature 
gradient at vanishing current, $S=-\Delta V/\Delta T|_{I=0}$, in the limit 
of $\Delta T\rightarrow 0$. The electric current is obtained within a 
single-particle picture using the Landauer approach 
\begin{equation}
I = \frac{e}{\pi \hbar} \int_{-\infty} ^\infty \mathcal{T}(E)(f_L(E)-f_R(E))dE\,\,\,,
\end{equation}
where  $\mathcal{T}(E)$ is the transmission coefficient, and $f_{R,L}$ are 
the Fermi distributions of the right and left leads.  
The Seebeck coefficient is calculated in the linear response regime, 
i.e. $|\Delta T|<<T$ and $|e\Delta V|<<\mu$, with $\mu$ being the equilibrium
chemical potential at the temperature $T$. 
It is given by\cite{cutler1969observation}
\begin{equation}
S(\mu,T)=\frac{1}{eT}\frac{\int_{-\infty} ^\infty \left(-\frac{\partial f}
{\partial E}\right) (E-\mu) \mathcal{T}(E) \, dE}
{ \int_{-\infty}^\infty \left(-\frac{\partial f}{\partial E}\right)  \mathcal{T}(E) 
\, dE}\,\,\,.
\label{Eq:S}
\end{equation}

\subsection{\label{sec:Analytical} Analytical Approach}
Before discussing further the trends observed in different 
regimes, we elaborate on analytical expansions in series for 
the electronic transmission and the thermoelectric coefficient.

The total number $N$ of propagating modes depends on the ribbon width $W$,
and the electronic transmission of a conducting channel $n$ can be formulated analytically.\cite{brey2006electronic,tworzydlo2006sub}
By applying the edge conditions of a metallic armchair ribbon to the Dirac 
equation, i.e. imposing that the wave functions values are zero in $ y = 0 $ 
and in $ y = W $. The transmission probability per channel is given by
\begin{equation}
\mathcal{T}_n = \frac{1}{\cosh ^2 \left( \pi n L / W \right)}, \,\, n = 0,\,1,\,2\,...
\end{equation}  
\noindent and the electronic conductance $G$ can be written as\footnote{There may be a discrepancy with previous works by a factor of 2 due to the spin freedom degree.}
\begin{equation}
G = \left(\frac{2 e^2}{h}\right) \sum ^{N-1}_{n=0} \mathcal{T}_n.
\label{exp_G}
\end{equation}

\begin{figure}[t!]
\includegraphics[clip,width=0.48\textwidth,angle=0,clip]{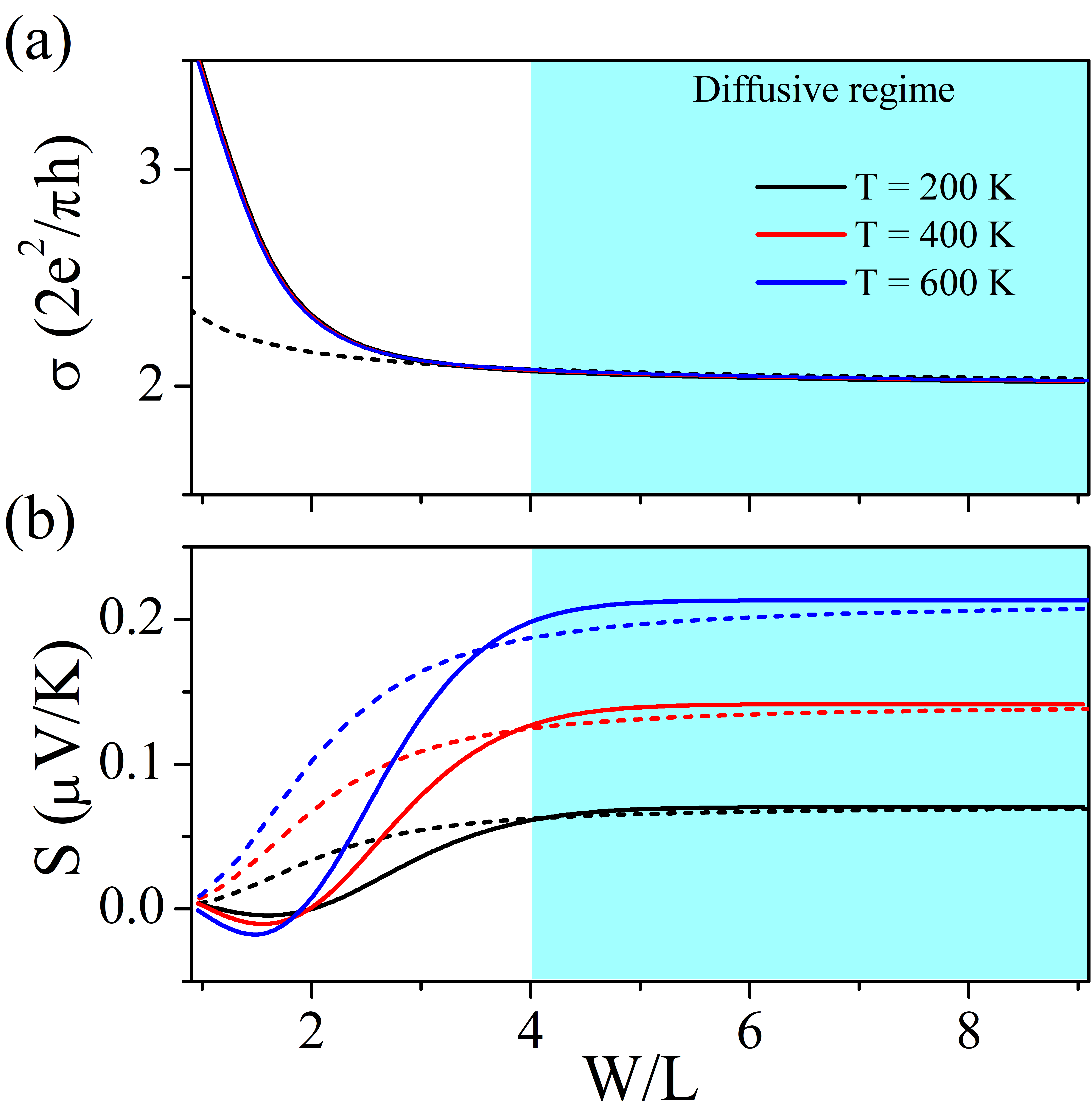}
\caption{(Color on-line) Conductivity $\sigma $ and Seebeck coefficient 
$S$ at the conductor Dirac point, as a function of the aspect ratio $W/L$ 
for different temperatures. We fix the length to $L = 3 \times 3 a_{CC}$ 
and vary the width $W$ only selecting the metallic armchair ribbons. 
Solid lines correspond to the numerical solutions of Eqs. \ref{LandauerG} 
and \ref{Eq:S}. Dashed lines correspond to the analytical results from the 
series expansion in Eqs. \ref{exp_G} and \ref{casilim}.}
\label{Fig:VC_WL}
\end{figure}


The Seebeck coefficient coefficient can be expressed using a Sommerfeld 
expansion for temperatures below the Fermi temperature.
\cite{abrahams1979scaling,sivan1986multichannel,sanchez2013scattering}
The Seebeck coefficient can be thus written as a ratio between the 
integrals $K_0$ and $K_1$, defined as
\begin{eqnarray}
K_0\!&=&\!\mathcal{T}+\frac{\pi^2}{6}\mathcal{T}^{(2)}\xi^2+\frac{7\pi^4}{360}
\mathcal{T}^{(4)}\xi^4+ O(\xi^6), \label{K0} \\
K_1\!&=& \!\frac{\pi^2}{3}\mathcal{T}^{(1)}\xi^2+ 
\frac{7\pi^4}{90}\mathcal{T}^{(3)}\xi^4+O(\xi^6),\label{K1}  
\end{eqnarray} \label{kas}
\noindent where $\xi=k_B T$ and 
$\mathcal{T}^{(n)}\equiv\mathcal{T}^{(n)}(\mu)=(d^n\mathcal{T}/dE^n)(\mu)$. 
When considering the lowest order, the Seebeck coefficient can be shorten as
\begin{eqnarray}
S\!&=&\!\frac{1}{eT} \frac{K_1}{K_0} = \frac{T \left(\pi k_B \right)^2 }{3e} 
\frac{\partial}{\partial E} \left[ \ln\left(K_0\right)\right]_{\mu}, \nonumber \\
 \!&\approx&\!\frac{T \left(\pi k_B \right)^2 }{3e} \frac{1}{\mathcal{T}} \frac{\partial 
 \mathcal{T}}{\partial E}\bigg|_{E = E_F} . 
\label{casilim}
\end{eqnarray}
\noindent A similar expression was found by solving the Boltzmann equation 
on a gentle temperature gradient.\cite{Parravicini:2000}
Finally, because the transmission $\mathcal{T}$ becomes smooth with the 
chemical potential near the Fermi level (see below 
Fig. \ref{Fig:VC_N35xEnergy_s}(a) and related comments), the Seebeck 
coefficient limit $S_\infty$ is written as
\begin{equation}
S\left(W/L \rightarrow\infty \right) = S_\infty = \frac{\left(\pi k_B\right)^2 }{3e} T. \label{S_limit}
\end{equation}

\section{\label{sec:discussion} Discussion}
To discuss the pseudodiffusive regimen, we have considered a tunnel 
junction configuration\cite{cresti2013impact}, in which the system 
between electrodes consists of a wide metallic nanoribbon parametrized 
by the length $L$ and the width  $W (>>L)$. 
Our discussion focuses first on the $W/L\rightarrow\infty$ limit 
for the electronic conductivity $\sigma$ and the Seebeck coefficient $S$.

The conductivity $\sigma$ dependence on the aspect ratio $W/L$ is 
shown in Fig. \ref{Fig:VC_WL}(a). We find that the conductivity as 
a function of temperature undergoes small modifications. Our results for a fix temperature are consistent with the existing literature.\cite{borunda2013ballistic,tworzydlo2006sub,cresti2013impact}
When $W/L\rightarrow\infty$, the conductivity tends to a constant 
value, namely $\sigma_{\infty} = 4e^2/\pi h$. Note that using the 
same conditions, it was shown that the limit for the Fano factor 
is also a constant 
$\mathcal{F}\rightarrow 1/3$.\cite{tworzydlo2006sub,danneau2008shot}
These findings are, therefore, the fingerprint of reaching a 
diffusive regime. We find that in practice, the pseudodiffusive limit 
is already reached when having $W/L > 4$, identified as the shaded 
region in Fig. \ref{Fig:VC_WL} when the curves are saturated. 
Furthermore, in the pseudodiffusive regime obtained by increasing 
the $W/L$ ratio, the numerical calculations using the Landauer formula 
(Eq. \ref{LandauerG}) approach become closer to the analytical expansion 
of the conductivity given by Eq. \ref{exp_G}.
These results agree with the electronic transport experiments, which do 
not show temperature dependence in the conductivity of 2D graphene 
with temperatures below $300$ K.\cite{hwang2007carrier}

Figure \ref{Fig:VC_WL}(b) shows the different behavior of the Seebeck 
coefficient for graphene nanoribbons when the aspect ratio $W/L$ changes. 
In the ballistic limit, for small $W/L$, the values of Seebeck coefficient 
have a minimum. When the temperature increases, the $W/L$ ratio at which $S$ 
has a minimum decreases, while the $S_{min}$ value increases.


The thermopower or Seebeck coefficient $S$, as defined in Eq. \ref{Eq:S} through the transmission probability, depends directly on the temperature 
and indirectly on the aspect ratio $W/L$. 
In Fig. \ref{Fig:VC_WL}(b) we exposed these dependences. Similarly 
to conductivity, it is possible to identify the pseudodiffusive 
regime ($W/L > 4$) when the Seebeck coefficient becomes saturated. 
Our numerical results show that the Seebeck coefficient is saturated 
as the $W/L$ ratio increases, and the $S_\infty$ limit depends on the temperature, following Eq. \ref{S_limit}.
Because the used Sommerfeld expansion is valid for temperatures below the Fermi temperature, the analytical and numerical results differs slightly 
as the temperature increases.
It is noteworthy that the experimental measurements estimated large 
values for the Fermi temperature as $T_F \sim 2490$ K in patterned 
epitaxial graphene\cite{berger2006electronic}, and  $T_F \sim 1300$ K 
for free standing 2D graphene being doped $n<10^{12} \text{cm}^{-2}$.\cite{hwang2007carrier} 

In the pseudodiffusive regimen, the Seebeck coefficient saturates 
because the curves converge asymptotically against the $W/L$ ratio. 
For large $W/L$ ratios, the Seebeck coefficient is highly dependent 
on the temperature, a result in agreement with the semi-classical 
Mott formula but not expected for a ballistic system.
In addition to the well-known limits $\sigma \rightarrow 4e^2/\pi h$ 
and $\mathcal{F} \rightarrow 1/3$, our results reveals other trend 
for electronic transport in the diffusive regime, that is,  
$S \rightarrow \frac{\left(\pi k_B\right)^2 }{3e} T$ as
defined in Eq. \ref{S_limit},  which shows the a linear dependence on temperature of the Seebeck coefficient coefficient.

\begin{figure}[ht!]
\includegraphics[clip,width=0.48\textwidth,angle=0,clip]{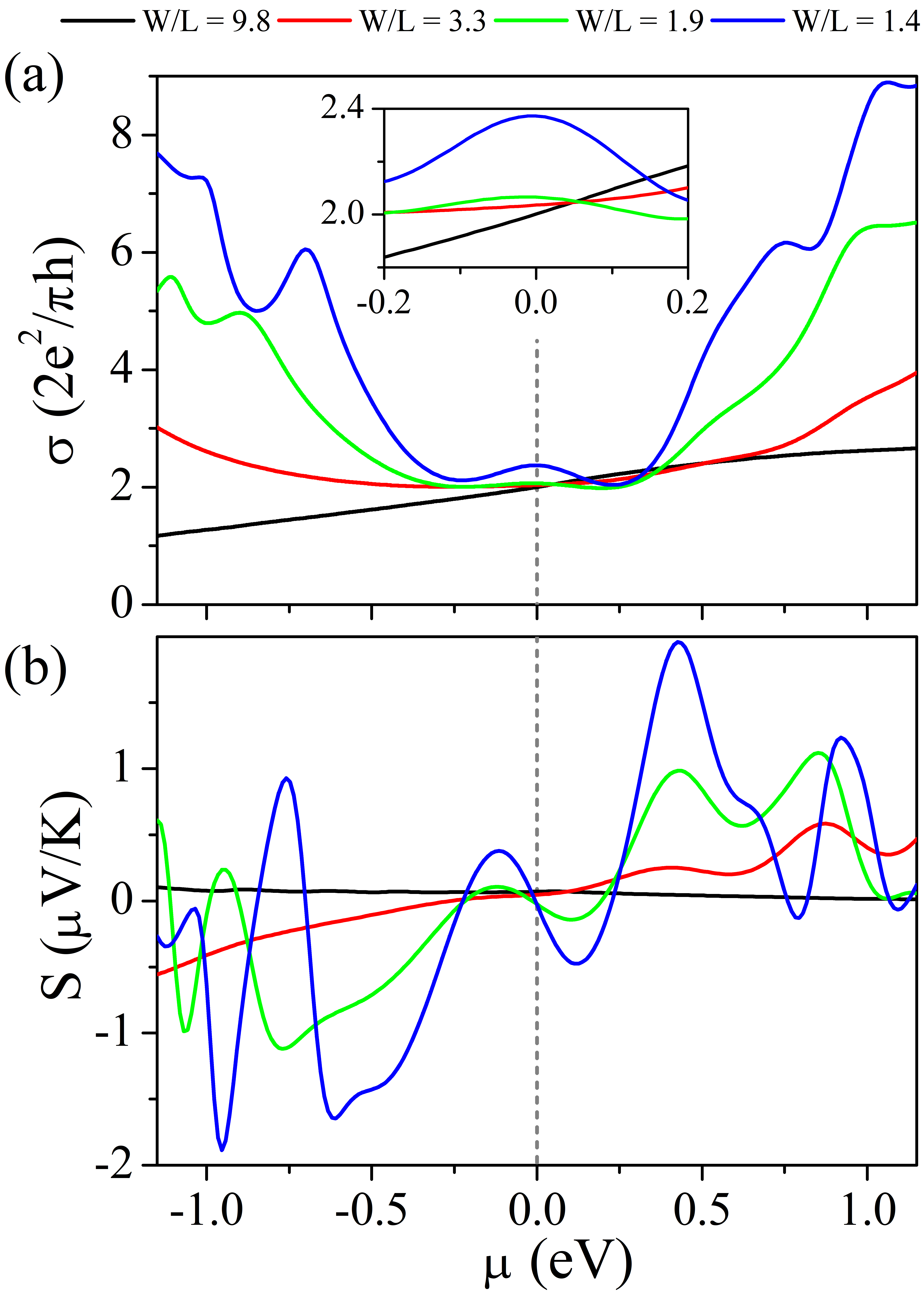}
\caption{(Color on-line) Conductivity and Seebeck coefficient as a 
function of the chemical potential for aGNRs with fix temperature 
$T = 200$ K  and fix ribbon width $N = 35$ $\left( W = 35 \sqrt{3} a_{CC} \right)$ and several $W/L$ ratios. }
\label{Fig:VC_N35xEnergy_s}
\end{figure}

Finally, we comment on the transition between ballistic and pseudodiffusive regimes looking at numerical calculations of conductivity and Seebeck coefficient versus  chemical potential. 
Figure \ref{Fig:VC_N35xEnergy_s} displays the behavior of $\sigma$ and 
$S$ depending on the chemical potential $\mu$ for different aspect 
ratios $W/L$, using a ribbon with $N = 35$ and a fix temperature of 
$T = 200$ K. 
In general, the conductivity decreases as the aspect ratio increases. 
In the pseudodiffusive region, for example with $W/L \sim 10$, the conductivity is linear with the chemical potential. This trend is 
already indicated in  a smaller range of $W/L$  ratios looking at a 
narrow energy window, as shown in the inset for the case of 
$W / L \sim 3$. The Seebeck coefficient $S$ in the ballistic regimen 
around the Fermi level is characterized by a maximum followed by a 
minimum. However, when the aspect ratio increases in the pseudodiffusive 
regime, the local maximum values decrease and expand reaching an almost constant value over the entire energy range around the Fermi energy.

\section{Final Remarks}
In summary we reproduce the well-known limits 
$\sigma \rightarrow 4e^2/\pi h$ and  $\mathcal{F} \rightarrow 1/3$ 
in the diffusive regime. Our results on Seebeck coefficient 
(or thermopower) allow us to add new features to the electronic transport 
of graphene nanoribbons in the diffusive regime. 
We find that $S \rightarrow \frac{\left(\pi k_B\right)^2 }{3e} T$ as 
defined in Eq. \ref{S_limit} so that the Seebeck coefficient depends 
linearly on the temperature.
In addition, we have performed numerical calculations, which in the asymptotic pseudodiffusive limit are in good agreement with the 
analytical expressions. 
Notice also that although we are dealing with wide and short metallic 
tunnel junctions, our results are following the estimates using the 
Boltzman equation for large and disordered systems, a fact which 
fulfills the semi-classical Mott formula.

\section*{Acknowledgments}
This work has been supported by the Project FIS2013-48286-C2-1-P 
of the Spanish Ministry of Economy and Competitiveness MINECO (JWG, AA). 
The Basque Government through the NANOMATERIALS project (Grant  IE14-393)
under the ETORTEK Program {\it Nanogune14}, and the University of the 
Basque Country (Grant No. IT-366-07). LR acknowledge support from the 
Fondecyt (Grants 1140388 and 1180914). JWG and AA acknowledge the hospitality of the Universidad T\'ecnica Federico Santa Mar\'ia.


\begin{thebibliography}{35}
\expandafter\ifx\csname natexlab\endcsname\relax\def\natexlab#1{#1}\fi
\expandafter\ifx\csname bibnamefont\endcsname\relax
  \def\bibnamefont#1{#1}\fi
\expandafter\ifx\csname bibfnamefont\endcsname\relax
  \def\bibfnamefont#1{#1}\fi
\expandafter\ifx\csname citenamefont\endcsname\relax
  \def\citenamefont#1{#1}\fi
\expandafter\ifx\csname url\endcsname\relax
  \def\url#1{\texttt{#1}}\fi
\expandafter\ifx\csname urlprefix\endcsname\relax\def\urlprefix{URL }\fi
\providecommand{\bibinfo}[2]{#2}
\providecommand{\eprint}[2][]{\url{#2}}

\bibitem[{\citenamefont{Tassin et~al.}(2013)\citenamefont{Tassin, Koschny, and
  Soukoulis}}]{tassin2013graphene}
\bibinfo{author}{\bibfnamefont{P.}~\bibnamefont{Tassin}},
  \bibinfo{author}{\bibfnamefont{T.}~\bibnamefont{Koschny}}, \bibnamefont{and}
  \bibinfo{author}{\bibfnamefont{C.~M.} \bibnamefont{Soukoulis}},
  \bibinfo{journal}{Science} \textbf{\bibinfo{volume}{341}},
  \bibinfo{pages}{620} (\bibinfo{year}{2013}).

\bibitem[{\citenamefont{Gonz{\'a}lez et~al.}(2011)\citenamefont{Gonz{\'a}lez,
  Pacheco, Rosales, and Orellana}}]{gonzalez2011transport}
\bibinfo{author}{\bibfnamefont{J.~W.} \bibnamefont{Gonz{\'a}lez}},
  \bibinfo{author}{\bibfnamefont{M.}~\bibnamefont{Pacheco}},
  \bibinfo{author}{\bibfnamefont{L.}~\bibnamefont{Rosales}}, \bibnamefont{and}
  \bibinfo{author}{\bibfnamefont{P.}~\bibnamefont{Orellana}},
  \bibinfo{journal}{Physical Review B} \textbf{\bibinfo{volume}{83}},
  \bibinfo{pages}{155450} (\bibinfo{year}{2011}).

\bibitem[{\citenamefont{Miao et~al.}(2007)\citenamefont{Miao, Wijeratne, Zhang,
  Coskun, Bao, and Lau}}]{miao2007phase}
\bibinfo{author}{\bibfnamefont{F.}~\bibnamefont{Miao}},
  \bibinfo{author}{\bibfnamefont{S.}~\bibnamefont{Wijeratne}},
  \bibinfo{author}{\bibfnamefont{Y.}~\bibnamefont{Zhang}},
  \bibinfo{author}{\bibfnamefont{U.}~\bibnamefont{Coskun}},
  \bibinfo{author}{\bibfnamefont{W.}~\bibnamefont{Bao}}, \bibnamefont{and}
  \bibinfo{author}{\bibfnamefont{C.}~\bibnamefont{Lau}},
  \bibinfo{journal}{Science} \textbf{\bibinfo{volume}{317}},
  \bibinfo{pages}{1530} (\bibinfo{year}{2007}).

\bibitem[{\citenamefont{Tan et~al.}(2007)\citenamefont{Tan, Zhang, Bolotin,
  Zhao, Adam, Hwang, Sarma, Stormer, and Kim}}]{tan2007measurement}
\bibinfo{author}{\bibfnamefont{Y.-W.} \bibnamefont{Tan}},
  \bibinfo{author}{\bibfnamefont{Y.}~\bibnamefont{Zhang}},
  \bibinfo{author}{\bibfnamefont{K.}~\bibnamefont{Bolotin}},
  \bibinfo{author}{\bibfnamefont{Y.}~\bibnamefont{Zhao}},
  \bibinfo{author}{\bibfnamefont{S.}~\bibnamefont{Adam}},
  \bibinfo{author}{\bibfnamefont{E.}~\bibnamefont{Hwang}},
  \bibinfo{author}{\bibfnamefont{S.~D.} \bibnamefont{Sarma}},
  \bibinfo{author}{\bibfnamefont{H.}~\bibnamefont{Stormer}}, \bibnamefont{and}
  \bibinfo{author}{\bibfnamefont{P.}~\bibnamefont{Kim}},
  \bibinfo{journal}{Physical review letters} \textbf{\bibinfo{volume}{99}},
  \bibinfo{pages}{246803} (\bibinfo{year}{2007}).

\bibitem[{\citenamefont{Chen et~al.}(2008)\citenamefont{Chen, Jang, Adam,
  Fuhrer, Williams, and Ishigami}}]{chen2008charged}
\bibinfo{author}{\bibfnamefont{J.-H.} \bibnamefont{Chen}},
  \bibinfo{author}{\bibfnamefont{C.}~\bibnamefont{Jang}},
  \bibinfo{author}{\bibfnamefont{S.}~\bibnamefont{Adam}},
  \bibinfo{author}{\bibfnamefont{M.}~\bibnamefont{Fuhrer}},
  \bibinfo{author}{\bibfnamefont{E.}~\bibnamefont{Williams}}, \bibnamefont{and}
  \bibinfo{author}{\bibfnamefont{M.}~\bibnamefont{Ishigami}},
  \bibinfo{journal}{Nature Physics} \textbf{\bibinfo{volume}{4}},
  \bibinfo{pages}{377} (\bibinfo{year}{2008}).

\bibitem[{\citenamefont{Martin et~al.}(2008)\citenamefont{Martin, Akerman,
  Ulbricht, Lohmann, Smet, Von~Klitzing, and Yacoby}}]{martin2008observation}
\bibinfo{author}{\bibfnamefont{J.}~\bibnamefont{Martin}},
  \bibinfo{author}{\bibfnamefont{N.}~\bibnamefont{Akerman}},
  \bibinfo{author}{\bibfnamefont{G.}~\bibnamefont{Ulbricht}},
  \bibinfo{author}{\bibfnamefont{T.}~\bibnamefont{Lohmann}},
  \bibinfo{author}{\bibfnamefont{J.~v.} \bibnamefont{Smet}},
  \bibinfo{author}{\bibfnamefont{K.}~\bibnamefont{Von~Klitzing}},
  \bibnamefont{and} \bibinfo{author}{\bibfnamefont{A.}~\bibnamefont{Yacoby}},
  \bibinfo{journal}{Nature Physics} \textbf{\bibinfo{volume}{4}},
  \bibinfo{pages}{144} (\bibinfo{year}{2008}).

\bibitem[{\citenamefont{Xia et~al.}(2009)\citenamefont{Xia, Chen, Li, and
  Tao}}]{xia2009measurement}
\bibinfo{author}{\bibfnamefont{J.}~\bibnamefont{Xia}},
  \bibinfo{author}{\bibfnamefont{F.}~\bibnamefont{Chen}},
  \bibinfo{author}{\bibfnamefont{J.}~\bibnamefont{Li}}, \bibnamefont{and}
  \bibinfo{author}{\bibfnamefont{N.}~\bibnamefont{Tao}},
  \bibinfo{journal}{Nature nanotechnology} \textbf{\bibinfo{volume}{4}},
  \bibinfo{pages}{505} (\bibinfo{year}{2009}).

\bibitem[{\citenamefont{Adam et~al.}(2007)\citenamefont{Adam, Hwang, Galitski,
  and Sarma}}]{adam2007self}
\bibinfo{author}{\bibfnamefont{S.}~\bibnamefont{Adam}},
  \bibinfo{author}{\bibfnamefont{E.}~\bibnamefont{Hwang}},
  \bibinfo{author}{\bibfnamefont{V.}~\bibnamefont{Galitski}}, \bibnamefont{and}
  \bibinfo{author}{\bibfnamefont{S.~D.} \bibnamefont{Sarma}},
  \bibinfo{journal}{Proceedings of the National Academy of Sciences}
  \textbf{\bibinfo{volume}{104}}, \bibinfo{pages}{18392}
  (\bibinfo{year}{2007}).

\bibitem[{\citenamefont{Nomura and MacDonald}(2007)}]{nomura2007quantum}
\bibinfo{author}{\bibfnamefont{K.}~\bibnamefont{Nomura}} \bibnamefont{and}
  \bibinfo{author}{\bibfnamefont{A.}~\bibnamefont{MacDonald}},
  \bibinfo{journal}{Physical review letters} \textbf{\bibinfo{volume}{98}},
  \bibinfo{pages}{076602} (\bibinfo{year}{2007}).

\bibitem[{\citenamefont{Tworzyd{\l}o et~al.}(2006)\citenamefont{Tworzyd{\l}o,
  Trauzettel, Titov, Rycerz, and Beenakker}}]{tworzydlo2006sub}
\bibinfo{author}{\bibfnamefont{J.}~\bibnamefont{Tworzyd{\l}o}},
  \bibinfo{author}{\bibfnamefont{B.}~\bibnamefont{Trauzettel}},
  \bibinfo{author}{\bibfnamefont{M.}~\bibnamefont{Titov}},
  \bibinfo{author}{\bibfnamefont{A.}~\bibnamefont{Rycerz}}, \bibnamefont{and}
  \bibinfo{author}{\bibfnamefont{C.~W.} \bibnamefont{Beenakker}},
  \bibinfo{journal}{Physical Review Letters} \textbf{\bibinfo{volume}{96}},
  \bibinfo{pages}{246802} (\bibinfo{year}{2006}).

\bibitem[{\citenamefont{Haskins et~al.}(2011)\citenamefont{Haskins,
  K{\i}nac{\i}, Sevik, Sevin{\c{c}}li, Cuniberti, and
  Çağ{\i}n}}]{haskins2011control}
\bibinfo{author}{\bibfnamefont{J.}~\bibnamefont{Haskins}},
  \bibinfo{author}{\bibfnamefont{A.}~\bibnamefont{K{\i}nac{\i}}},
  \bibinfo{author}{\bibfnamefont{C.}~\bibnamefont{Sevik}},
  \bibinfo{author}{\bibfnamefont{H.}~\bibnamefont{Sevin{\c{c}}li}},
  \bibinfo{author}{\bibfnamefont{G.}~\bibnamefont{Cuniberti}},
  \bibnamefont{and}
  \bibinfo{author}{\bibfnamefont{T.}~\bibnamefont{Çağ{\i}n}},
  \bibinfo{journal}{ACS nano} \textbf{\bibinfo{volume}{5}},
  \bibinfo{pages}{3779} (\bibinfo{year}{2011}).

\bibitem[{\citenamefont{Zuev et~al.}(2009)\citenamefont{Zuev, Chang, and
  Kim}}]{zuev2009thermoelectric}
\bibinfo{author}{\bibfnamefont{Y.~M.} \bibnamefont{Zuev}},
  \bibinfo{author}{\bibfnamefont{W.}~\bibnamefont{Chang}}, \bibnamefont{and}
  \bibinfo{author}{\bibfnamefont{P.}~\bibnamefont{Kim}},
  \bibinfo{journal}{Physical Review Letters} \textbf{\bibinfo{volume}{102}},
  \bibinfo{pages}{096807} (\bibinfo{year}{2009}).

\bibitem[{\citenamefont{Wei et~al.}(2009)\citenamefont{Wei, Bao, Pu, Lau, and
  Shi}}]{wei2009anomalous}
\bibinfo{author}{\bibfnamefont{P.}~\bibnamefont{Wei}},
  \bibinfo{author}{\bibfnamefont{W.}~\bibnamefont{Bao}},
  \bibinfo{author}{\bibfnamefont{Y.}~\bibnamefont{Pu}},
  \bibinfo{author}{\bibfnamefont{C.~N.} \bibnamefont{Lau}}, \bibnamefont{and}
  \bibinfo{author}{\bibfnamefont{J.}~\bibnamefont{Shi}},
  \bibinfo{journal}{Physical review letters} \textbf{\bibinfo{volume}{102}},
  \bibinfo{pages}{166808} (\bibinfo{year}{2009}).

\bibitem[{\citenamefont{Babichev et~al.}(2013)\citenamefont{Babichev,
  Gasumyants, and Butko}}]{babichev2013resistivity}
\bibinfo{author}{\bibfnamefont{A.~V.} \bibnamefont{Babichev}},
  \bibinfo{author}{\bibfnamefont{V.~E.} \bibnamefont{Gasumyants}},
  \bibnamefont{and} \bibinfo{author}{\bibfnamefont{V.~Y.} \bibnamefont{Butko}},
  \bibinfo{journal}{Journal of Applied Physics} \textbf{\bibinfo{volume}{113}},
  \bibinfo{pages}{076101} (\bibinfo{year}{2013}).

\bibitem[{\citenamefont{Xiao et~al.}(2011)\citenamefont{Xiao, Dong, Song, Liu,
  Tay, Wu, Li, Zhao, Yu, Zhang et~al.}}]{xiao2011enhanced}
\bibinfo{author}{\bibfnamefont{N.}~\bibnamefont{Xiao}},
  \bibinfo{author}{\bibfnamefont{X.}~\bibnamefont{Dong}},
  \bibinfo{author}{\bibfnamefont{L.}~\bibnamefont{Song}},
  \bibinfo{author}{\bibfnamefont{D.}~\bibnamefont{Liu}},
  \bibinfo{author}{\bibfnamefont{Y.}~\bibnamefont{Tay}},
  \bibinfo{author}{\bibfnamefont{S.}~\bibnamefont{Wu}},
  \bibinfo{author}{\bibfnamefont{L.-J.} \bibnamefont{Li}},
  \bibinfo{author}{\bibfnamefont{Y.}~\bibnamefont{Zhao}},
  \bibinfo{author}{\bibfnamefont{T.}~\bibnamefont{Yu}},
  \bibinfo{author}{\bibfnamefont{H.}~\bibnamefont{Zhang}},
  \bibnamefont{et~al.}, \bibinfo{journal}{Acs Nano}
  \textbf{\bibinfo{volume}{5}}, \bibinfo{pages}{2749} (\bibinfo{year}{2011}).

\bibitem[{\citenamefont{Sankeshwar et~al.}(2013)\citenamefont{Sankeshwar,
  Kubakaddi, and Mulimani}}]{sankeshwar2013thermoelectric}
\bibinfo{author}{\bibfnamefont{N.}~\bibnamefont{Sankeshwar}},
  \bibinfo{author}{\bibfnamefont{S.}~\bibnamefont{Kubakaddi}},
  \bibnamefont{and} \bibinfo{author}{\bibfnamefont{B.}~\bibnamefont{Mulimani}},
  \bibinfo{journal}{Advances in Graphene Science}
  \textbf{\bibinfo{volume}{10}}, \bibinfo{pages}{56720} (\bibinfo{year}{2013}).

\bibitem[{\citenamefont{Borunda et~al.}(2013)\citenamefont{Borunda, Hennig, and
  Heller}}]{borunda2013ballistic}
\bibinfo{author}{\bibfnamefont{M.~F.} \bibnamefont{Borunda}},
  \bibinfo{author}{\bibfnamefont{H.}~\bibnamefont{Hennig}}, \bibnamefont{and}
  \bibinfo{author}{\bibfnamefont{E.~J.} \bibnamefont{Heller}},
  \bibinfo{journal}{Physical Review B} \textbf{\bibinfo{volume}{88}},
  \bibinfo{pages}{125415} (\bibinfo{year}{2013}).

\bibitem[{\citenamefont{Cresti et~al.}(2013)\citenamefont{Cresti, Louvet,
  Ortmann, Van~Tuan, Lenarczyk, Huhs, and Roche}}]{cresti2013impact}
\bibinfo{author}{\bibfnamefont{A.}~\bibnamefont{Cresti}},
  \bibinfo{author}{\bibfnamefont{T.}~\bibnamefont{Louvet}},
  \bibinfo{author}{\bibfnamefont{F.}~\bibnamefont{Ortmann}},
  \bibinfo{author}{\bibfnamefont{D.}~\bibnamefont{Van~Tuan}},
  \bibinfo{author}{\bibfnamefont{P.}~\bibnamefont{Lenarczyk}},
  \bibinfo{author}{\bibfnamefont{G.}~\bibnamefont{Huhs}}, \bibnamefont{and}
  \bibinfo{author}{\bibfnamefont{S.}~\bibnamefont{Roche}},
  \bibinfo{journal}{Crystals} \textbf{\bibinfo{volume}{3}},
  \bibinfo{pages}{289} (\bibinfo{year}{2013}).

\bibitem[{\citenamefont{Dietl et~al.}(2009)\citenamefont{Dietl, Metalidis,
  Golubev, San-Jose, Prada, Schomerus, and Sch{\"o}n}}]{dietl2009disorder}
\bibinfo{author}{\bibfnamefont{P.}~\bibnamefont{Dietl}},
  \bibinfo{author}{\bibfnamefont{G.}~\bibnamefont{Metalidis}},
  \bibinfo{author}{\bibfnamefont{D.}~\bibnamefont{Golubev}},
  \bibinfo{author}{\bibfnamefont{P.}~\bibnamefont{San-Jose}},
  \bibinfo{author}{\bibfnamefont{E.}~\bibnamefont{Prada}},
  \bibinfo{author}{\bibfnamefont{H.}~\bibnamefont{Schomerus}},
  \bibnamefont{and}
  \bibinfo{author}{\bibfnamefont{G.}~\bibnamefont{Sch{\"o}n}},
  \bibinfo{journal}{Physical Review B} \textbf{\bibinfo{volume}{79}},
  \bibinfo{pages}{195413} (\bibinfo{year}{2009}).

\bibitem[{\citenamefont{Nardelli}(1999)}]{nardelli1999electronic}
\bibinfo{author}{\bibfnamefont{M.~B.} \bibnamefont{Nardelli}},
  \bibinfo{journal}{Physical Review B} \textbf{\bibinfo{volume}{60}},
  \bibinfo{pages}{7828} (\bibinfo{year}{1999}).

\bibitem[{\citenamefont{Datta}(1997)}]{datta1997electronic}
\bibinfo{author}{\bibfnamefont{S.}~\bibnamefont{Datta}},
  \emph{\bibinfo{title}{Electronic Transport in Mesoscopic Systems}}
  (\bibinfo{publisher}{Cambridge university press}, \bibinfo{year}{1997}).

\bibitem[{\citenamefont{Gonz{\'a}lez et~al.}(2015)\citenamefont{Gonz{\'a}lez,
  Rosales, Pacheco, and Ayuela}}]{gonzalez2015electron}
\bibinfo{author}{\bibfnamefont{J.~W.} \bibnamefont{Gonz{\'a}lez}},
  \bibinfo{author}{\bibfnamefont{L.}~\bibnamefont{Rosales}},
  \bibinfo{author}{\bibfnamefont{M.}~\bibnamefont{Pacheco}}, \bibnamefont{and}
  \bibinfo{author}{\bibfnamefont{A.}~\bibnamefont{Ayuela}},
  \bibinfo{journal}{Physical Chemistry Chemical Physics}
  \textbf{\bibinfo{volume}{17}}, \bibinfo{pages}{24707} (\bibinfo{year}{2015}).

\bibitem[{\citenamefont{Saito et~al.}(2003)\citenamefont{Saito, Dresselhaus,
  and Dresselhaus}}]{Saito:2003}
\bibinfo{author}{\bibfnamefont{R.}~\bibnamefont{Saito}},
  \bibinfo{author}{\bibfnamefont{G.}~\bibnamefont{Dresselhaus}},
  \bibnamefont{and} \bibinfo{author}{\bibfnamefont{M.~S.}
  \bibnamefont{Dresselhaus}}, \emph{\bibinfo{title}{Physical Properties of
  Carbon Nanotubes}} (\bibinfo{publisher}{Imperial College Press},
  \bibinfo{address}{London, UK}, \bibinfo{year}{2003}).

\bibitem[{\citenamefont{Charlier et~al.}(2007)\citenamefont{Charlier, Blase,
  and Roche}}]{Charlier:2007}
\bibinfo{author}{\bibfnamefont{J.-C.} \bibnamefont{Charlier}},
  \bibinfo{author}{\bibfnamefont{X.}~\bibnamefont{Blase}}, \bibnamefont{and}
  \bibinfo{author}{\bibfnamefont{S.}~\bibnamefont{Roche}},
  \bibinfo{journal}{Reviews of Modern Physics} \textbf{\bibinfo{volume}{79}},
  \bibinfo{pages}{677} (\bibinfo{year}{2007}).

\bibitem[{\citenamefont{Laird et~al.}(2015)\citenamefont{Laird, Kuemmeth,
  Steele, Grove-Rasmussen, Nyg{\aa}rd, Flensberg, and
  Kouwenhoven}}]{Laird:2015}
\bibinfo{author}{\bibfnamefont{E.~A.} \bibnamefont{Laird}},
  \bibinfo{author}{\bibfnamefont{F.}~\bibnamefont{Kuemmeth}},
  \bibinfo{author}{\bibfnamefont{G.~A.} \bibnamefont{Steele}},
  \bibinfo{author}{\bibfnamefont{K.}~\bibnamefont{Grove-Rasmussen}},
  \bibinfo{author}{\bibfnamefont{J.}~\bibnamefont{Nyg{\aa}rd}},
  \bibinfo{author}{\bibfnamefont{K.}~\bibnamefont{Flensberg}},
  \bibnamefont{and} \bibinfo{author}{\bibfnamefont{L.~P.}
  \bibnamefont{Kouwenhoven}}, \bibinfo{journal}{Reviews Of Modern Physics}
  \textbf{\bibinfo{volume}{87}}, \bibinfo{pages}{703} (\bibinfo{year}{2015}).

\bibitem[{\citenamefont{Gonz{\'a}lez et~al.}(2010)\citenamefont{Gonz{\'a}lez,
  Pacheco, Rosales, and Orellana}}]{gonzalez2010bound}
\bibinfo{author}{\bibfnamefont{J.~W.} \bibnamefont{Gonz{\'a}lez}},
  \bibinfo{author}{\bibfnamefont{M.}~\bibnamefont{Pacheco}},
  \bibinfo{author}{\bibfnamefont{L.}~\bibnamefont{Rosales}}, \bibnamefont{and}
  \bibinfo{author}{\bibfnamefont{P.}~\bibnamefont{Orellana}},
  \bibinfo{journal}{EPL (Europhysics Letters)} \textbf{\bibinfo{volume}{91}},
  \bibinfo{pages}{66001} (\bibinfo{year}{2010}).

\bibitem[{\citenamefont{Cutler and Mott}(1969)}]{cutler1969observation}
\bibinfo{author}{\bibfnamefont{M.}~\bibnamefont{Cutler}} \bibnamefont{and}
  \bibinfo{author}{\bibfnamefont{N.~F.} \bibnamefont{Mott}},
  \bibinfo{journal}{Physical Review} \textbf{\bibinfo{volume}{181}},
  \bibinfo{pages}{1336} (\bibinfo{year}{1969}).

\bibitem[{\citenamefont{Brey and Fertig}(2006)}]{brey2006electronic}
\bibinfo{author}{\bibfnamefont{L.}~\bibnamefont{Brey}} \bibnamefont{and}
  \bibinfo{author}{\bibfnamefont{H.}~\bibnamefont{Fertig}},
  \bibinfo{journal}{Physical Review B} \textbf{\bibinfo{volume}{73}},
  \bibinfo{pages}{235411} (\bibinfo{year}{2006}).

\bibitem[{\citenamefont{Abrahams et~al.}(1979)\citenamefont{Abrahams, Anderson,
  Licciardello, and Ramakrishnan}}]{abrahams1979scaling}
\bibinfo{author}{\bibfnamefont{E.}~\bibnamefont{Abrahams}},
  \bibinfo{author}{\bibfnamefont{P.}~\bibnamefont{Anderson}},
  \bibinfo{author}{\bibfnamefont{D.}~\bibnamefont{Licciardello}},
  \bibnamefont{and}
  \bibinfo{author}{\bibfnamefont{T.}~\bibnamefont{Ramakrishnan}},
  \bibinfo{journal}{Physical Review Letters} \textbf{\bibinfo{volume}{42}},
  \bibinfo{pages}{673} (\bibinfo{year}{1979}).

\bibitem[{\citenamefont{Sivan and Imry}(1986)}]{sivan1986multichannel}
\bibinfo{author}{\bibfnamefont{U.}~\bibnamefont{Sivan}} \bibnamefont{and}
  \bibinfo{author}{\bibfnamefont{Y.}~\bibnamefont{Imry}},
  \bibinfo{journal}{Physical Review B} \textbf{\bibinfo{volume}{33}},
  \bibinfo{pages}{551} (\bibinfo{year}{1986}).

\bibitem[{\citenamefont{S{\'a}nchez and
  L{\'o}pez}(2013)}]{sanchez2013scattering}
\bibinfo{author}{\bibfnamefont{D.}~\bibnamefont{S{\'a}nchez}} \bibnamefont{and}
  \bibinfo{author}{\bibfnamefont{R.}~\bibnamefont{L{\'o}pez}},
  \bibinfo{journal}{Physical Review Letters} \textbf{\bibinfo{volume}{110}},
  \bibinfo{pages}{026804} (\bibinfo{year}{2013}).

\bibitem[{\citenamefont{Grosso and Parravicini}(2000)}]{Parravicini:2000}
\bibinfo{author}{\bibfnamefont{G.}~\bibnamefont{Grosso}} \bibnamefont{and}
  \bibinfo{author}{\bibfnamefont{G.}~\bibnamefont{Parravicini}},
  \emph{\bibinfo{title}{Solid State Physics}} (\bibinfo{publisher}{Academic
  Press}, \bibinfo{address}{London, UK}, \bibinfo{year}{2000}).

\bibitem[{\citenamefont{Danneau et~al.}(2008)\citenamefont{Danneau, Wu,
  Craciun, Russo, Tomi, Salmilehto, Morpurgo, and Hakonen}}]{danneau2008shot}
\bibinfo{author}{\bibfnamefont{R.}~\bibnamefont{Danneau}},
  \bibinfo{author}{\bibfnamefont{F.}~\bibnamefont{Wu}},
  \bibinfo{author}{\bibfnamefont{M.}~\bibnamefont{Craciun}},
  \bibinfo{author}{\bibfnamefont{S.}~\bibnamefont{Russo}},
  \bibinfo{author}{\bibfnamefont{M.}~\bibnamefont{Tomi}},
  \bibinfo{author}{\bibfnamefont{J.}~\bibnamefont{Salmilehto}},
  \bibinfo{author}{\bibfnamefont{A.}~\bibnamefont{Morpurgo}}, \bibnamefont{and}
  \bibinfo{author}{\bibfnamefont{P.~J.} \bibnamefont{Hakonen}},
  \bibinfo{journal}{Physical review letters} \textbf{\bibinfo{volume}{100}},
  \bibinfo{pages}{196802} (\bibinfo{year}{2008}).

\bibitem[{\citenamefont{Hwang et~al.}(2007)\citenamefont{Hwang, Adam, and
  Sarma}}]{hwang2007carrier}
\bibinfo{author}{\bibfnamefont{E.}~\bibnamefont{Hwang}},
  \bibinfo{author}{\bibfnamefont{S.}~\bibnamefont{Adam}}, \bibnamefont{and}
  \bibinfo{author}{\bibfnamefont{S.~D.} \bibnamefont{Sarma}},
  \bibinfo{journal}{Physical review letters} \textbf{\bibinfo{volume}{98}},
  \bibinfo{pages}{186806} (\bibinfo{year}{2007}).

\bibitem[{\citenamefont{Berger et~al.}(2006)\citenamefont{Berger, Song, Li, Wu,
  Brown, Naud, Mayou, Li, Hass, Marchenkov et~al.}}]{berger2006electronic}
\bibinfo{author}{\bibfnamefont{C.}~\bibnamefont{Berger}},
  \bibinfo{author}{\bibfnamefont{Z.}~\bibnamefont{Song}},
  \bibinfo{author}{\bibfnamefont{X.}~\bibnamefont{Li}},
  \bibinfo{author}{\bibfnamefont{X.}~\bibnamefont{Wu}},
  \bibinfo{author}{\bibfnamefont{N.}~\bibnamefont{Brown}},
  \bibinfo{author}{\bibfnamefont{C.}~\bibnamefont{Naud}},
  \bibinfo{author}{\bibfnamefont{D.}~\bibnamefont{Mayou}},
  \bibinfo{author}{\bibfnamefont{T.}~\bibnamefont{Li}},
  \bibinfo{author}{\bibfnamefont{J.}~\bibnamefont{Hass}},
  \bibinfo{author}{\bibfnamefont{A.~N.} \bibnamefont{Marchenkov}},
  \bibnamefont{et~al.}, \bibinfo{journal}{Science}
  \textbf{\bibinfo{volume}{312}}, \bibinfo{pages}{1191} (\bibinfo{year}{2006}).

\end{thebibliography}

\end{document}